# Science and Facebook: the same popularity law!

Zoltán Néda[a], Levente Varga[a] and Tamás S. Biró[b]

[a] Babeş-Bolyai University, Dept. of Physics, Cluj-Napoca, Romania
[b] Wigner Research Center of The Hungarian Academy of Sciences, Budapest, Hungary
E-mail: zneda@phys.ubbcluj.ro

## Abstract

*The distribution of scientific citations for publications selected with different rules (author, topic, institution, country, journal, etc...) collapse on a single curve if one plots the citations relative to their mean value. We find that the distribution of "shares" for the Facebook posts rescale in the same manner to the very same curve with scientific citations. This finding suggests that citations are subjected to the same growth mechanism with Facebook popularity measures, being influenced by a statistically similar social environment and selection mechanism. In a simple master-equation approach the exponential growth of the number of publications and a preferential selection mechanism leads to a Tsallis-Pareto distribution offering an excellent description for the observed statistics. Based on our model and on the data derived from PubMed we predict that according to the present trend the average citations per scientific publications exponentially relaxes to about 4.*

**Classification**: Physical Sciences/Statistics and Sustainability Science

The number of citations for a publication is basically a social popularity measure for it, while it is considered to reflect the quality and impact of the research.
Citations are thus in our focus when evaluating researchers, groups and institutes [1-3]. The statistics and dynamics of citations are studied in several works [4-9] and lately we assisted to many serious debates on their use for quantifying objectively the quality and impact of a given research [1-3, 10-13]. In view of this, further scientific arguments or novel information regarding the citation statistics and its similarity to other social selection mechanisms is of enhanced importance.

It has been reported [4-6] that citations for scientific papers, selected according to an arbitrary collection rule, like author, topic, publication year, institution, journal, etc..., rescale on a common curve if considering their value relative to the average in the selected group. More specifically, if one computes for the selected set the probability density $f(x)$ for one paper to have $x$ citations, and represent graphically the $\langle x \rangle \cdot f(x)$ value as a function of $x/\langle x \rangle$, the probability distribution function (PDF) obtained for different sets will collapse on the same curve (see the figures in [4-6] and Figure 1). We denoted here by $\langle x \rangle$ the mean value of $x$, or the first moment of the PDF. For high citation numbers a clear power-law trend is visible, especially if one considers datasets where the $x/\langle x \rangle > 10$ domain is visible. There is, however, a very active debate on fitting this rescaled curve [4-6,9,14-19]. Researchers have suggested lognormal, negative binomial, Wakeby and power-law tailed distributions to fit the entire curve. Recent results [16,18,19] favour a Tsallis-Pareto (TP) [20,21] type hooked distribution, albeit the lognormal distribution is still in use [6]. The obvious scale-free nature of the tail and accordingly the observed invariance relative to mixing or selecting just a part of the ensemble is however a major argument in favour of the TP distribution.

Biology, physics and socio-economic phenomena offer many intriguing examples of scale-free distributions in complex systems [22-24]. The celebrated Zipf law [25], or many other power-law tailed distributions are widely known and well-studied [26]. The pure power-law, however, is not a distribution in the strict mathematical sense since it cannot be normalized for the whole interval between zero and infinity. Quite frequently we do not even have a large enough scaling interval to prove or disprove the presence of pure power-law distributions [27]. On the other hand the Tsallis-Pareto distribution [20,21]

$$f(x) = \frac{g}{(g-1)\langle x \rangle}\left(1 + \frac{x}{(g-1)\langle x \rangle}\right)^{-1-g} \quad (1)$$

is a proper probability density function (PDF) with a power-law like tail. It has been found that many heavy-tailed distributions are well fitted by using the above PDF [23]. Although this is not strictly a scale-free distribution, one can numerically check that for $g > 1$ exponents and for large enough $x/\langle x \rangle$ the scale free properties and invariance under mixing or splitting of the dataset are well satisfied.

A simple exercise on citation data collected from more than 750 000 ISI Web of Science (WOS) publications (mapping a part of the WOS citation network by using an Internet robot, please see the Methods section), draws the shape of this universal curve. If one makes a simple data processing exercise from the total number of citations received in ten years for all ISI indexed journals (InCites, Journal Citation Reports [28]), the data (JCR) rescale on the very same curve. If we select now data for the publications authored by one researcher, for the publications published in a given journal in one given year or by authors associated to a given institute, the data rescale again. For $x/\langle x \rangle \geq 0.1$ the collapsed data can be nicely fitted with a one-parameter TP PDF, using $g \approx 1.4$. (see Figure 1). As we already emphasized, this type of fit has the advantage that the scale-free property for $x/\langle x \rangle \geq 0.1$ is evident and also explains the invariance of the distribution when combining several data sets.

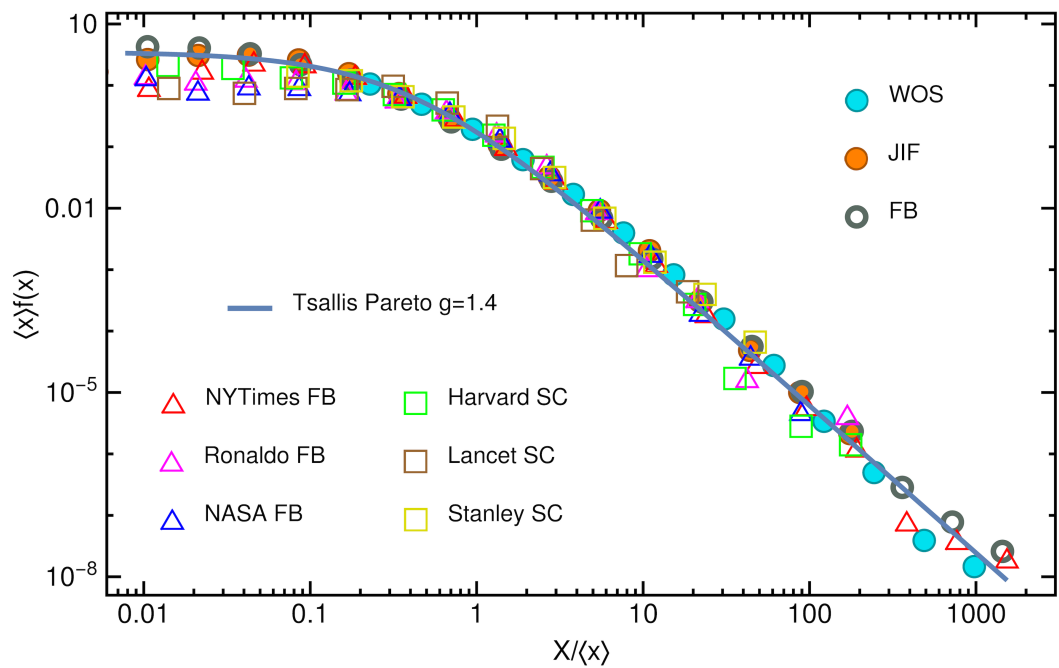

**Figure 1.** Rescaled distribution of the citation (share) numbers. $f(x)$ is the probability density (PDF) for one paper (post) to have $x$ citations/shares. We present the $\langle x \rangle \cdot f(x)$ value as a function of $x/\langle x \rangle$ ($\langle x \rangle$ the mean value, or first moment of the PDF). For high citation number a clear power-law trend is visible. Different symbols are for different datasets as illustrated in the legend. The considered datasets are described in the Methods section. For high $x/\langle x \rangle$ a clear power-law trend is visible. The entire curve can be well-fitted with a TP distribution (1) with $g \approx 1.4$.

A similar study can be performed on different Facebook pages for their posts (see the Methods section). Instead of citations the popularity proxy for a post is the number of "shares" it receives. "Share" is a stronger selection rule than the simple "like", and it's role is similar to citations in Science. Interestingly the PDF constructed for "shares" collected from 16 different Facebook users (in total more than 150 000 posts) rescale on the very same curve with the Scientific Citations (Figure 1). The universal TP distribution with $g \approx 1.4$ suggests a common growth mechanism for the Facebook Shares and Scientific Citations. Reducing now the Facebook data on users the rescaled PDF behaves in a similar manner (see the Methods section and for some selected data the plots on Figure 1). Due to the larger scatter for the data points resulting from the reduced data size (both for scientific citations and Facebook shares) we cannot conclude however the same Pareto exponent, just observe the similar trend. The invariance of the distributions relative to the splitting of the data is in agreement with the scale-free properties of this distribution.

**A simple stochastic growth model**

Many models have been already considered for explaining the dynamics of citations [29-31] and the observed universality in the rescaled PDF [32,33]. A simple explanation for this intriguing universality can be given by considering a simple coarse-grained master equation for the growth process and assuming an exponential growth of publications (post) number as a function of time together with a linear preferential growth rate in the flow (see Appendix 1).

The approach considered here is the simplest mean-field type approximation where only the stochastic nature of the growth process is taken into account and the specificity of the posts quality are coarse-grained. The exponential growth of the number of publications which are the carriers of the citations is known (see for example [34,35]). In a recent statement form Mark Zuckenberg we also learn that the information sharing activity on Facebook is also growing exponentially (see for example [36]).

On the other hand the linear preferential growth rate hypothesis or the commonly known Matthew effect ("For to all those who have, more will be given") has been highlighted in various social systems [37,38]. The presence of the Matthew effect in citation and science was also discussed in many previous publications [39,40]. In such manner the two main assumptions of our simple model are all reasonable, and can be applied both to Facebook posts and scientific articles. The Markov-like process constructed on these bases can be analytically solved also in the continuous limit where it leads to a TP (eq. 1) probability distribution (see Appendix 1). From the model we learn that the parameter $g$ in the TP distribution, governing the power-law tail, is just the ratio of the exponential growth rate $\gamma$ to the proportionality constant $\sigma$ for the linear preferential growth: $g = \gamma/\sigma$. The fact that the obtained $g$ value is independent from the way we construct the studied ensemble and it is the same for Facebook posts and Scientific Publications is intriguing. It can be understood by taking into account that both phenomena are taking place on a social network with similar topological properties, where the released information amount is increasing exponentially and the selection rules for its transmission are adapted to the increase rate.

From the promising fit displayed in Figure 1, using $g \approx 1.4$, we gain confidence in the statistical prediction capability of our simple mean-field type approximation. We elaborate thus further, on our model and make some statistical predictions on the expected evolution of the average number of citations (shares) per publication (post). Simple mathematics (see Appendix 2) will convince us that for $\gamma > \sigma$ ($g > 1$) the number of citations per article (shares per Facebook posts) exponentially relaxes to the equilibrium value implied by the TP distribution. We can also derive results for the time evolution of the yearly incoming total citation (shares) number, $c(t)$ (Appendix 2). For the case of scientific articles indexed in MEDLINE/PubMed (see the Methods section) we can determine the $\gamma \approx 0.06$ value (Figure 2a), which leads to $\sigma \approx 0.043$. A simple fitting exercise on the $c(t)$ curve using the data for PubMed, leads us to $b \approx 1.6$ (Figure 2a). According to these results we predict that the average number of citations per article (for the case of PubMed indexed articles) will relax to $b/(g-1) \approx 4$ (Figure 2b).

## Conclusions

Our **conclusions** are thus pretty clear: Science and Facebook show the same popularity pattern which can be simply understood by a coarse-grained master equation approach where we admit the exponentially increasing amount of information together with a "rich gets richer" preferential information filtering mechanism. Our model predicts that the average number of citations per publication (shares per Facebook posts) exponentially relaxes to a constant value, suggesting that our society does not inflate this popularity measure. For scientific articles we predict this number to be approximately 4.

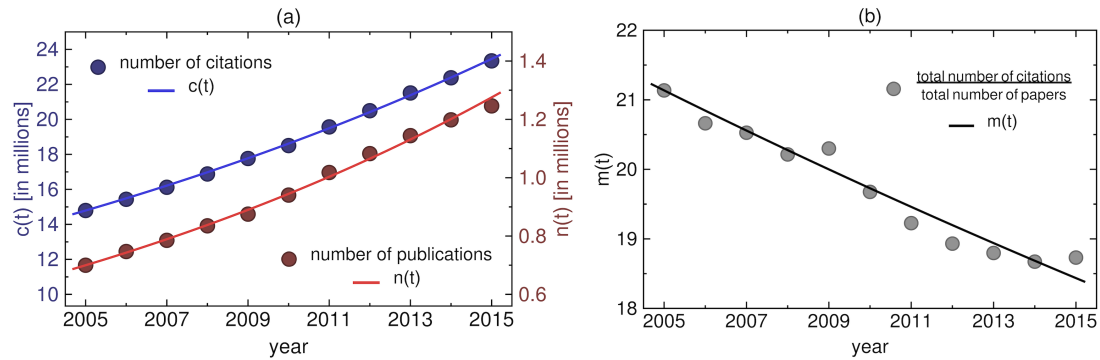

**Figure 2**. Results for the MEDLINE/PubMed database. **Figure 2a** illustrates the time evolution of the yearly indexed papers, $n(t)$, and the total number of citations, $c(t)$, introduced by them for each year in the 2005-2015 time interval. The trend $n(t)$ can be nicely fitted (red curve) with an exponential curve with $\gamma = 0.06$ using $t_0 = 2005$ and $n_0 = 699915$. Using $t_0 = 2005$, $n_0 = 699915$, $c_0 = 14792864$, $g = 1.4$ and $\gamma = 0.06$ ($\sigma = \gamma/g \approx 0.043$) the trend for $c(t)$ given by equation (2.3) can be fitted by choosing $b \approx 1.6$. **Figure 2b** illustrates the time evolution for the yearly incoming total number of citations divided by the total number of new papers, $m(t)$. Using the parameters from $n(t)$ and $c(t)$ the $m(t)$ trend given by equation (2.1) is plotted by the black curve.

## Methods

The data plotted in Figure 1 was collected as follows:

For the WOS dataset (Scientific Citations from ISI Web of Science) we used an Internet robot, that started form a given article and mapped all the papers that were cited by this. We have done this for a depth of four levels, and recorded the total number of citation for all ISI indexed articles that were reached with this procedure. In total more than 750 000 articles were mapped. For the JCR dataset we have downloaded the table from InCites, Journal Citation Reports [28], and recorded "total number of citations" for each of the (more than 12 000) indexed journals. For the reduced datasets we followed the methodology described in [6] selecting by random some Institutes, Journals and researchers. We extracted from ISI Web of Science the citations up to the present date for articles published in 1990 with authors from Harvard University. In the same manner for journals we selected papers published in The Lancet (Elsevier) in 1990 and recorded their citations up to the present date. Since our results were in agreement with the one published in [6], we concluded that the results for other Institutes and Journals rescale on the very same curve as it is illustrated in [6]. To complete the study on citation distribution with an even more challenging dataset we have selected a single author from physics (Prof. H. E. Stanley from the Boston University, USA) with an impressive number of publications (965 ISI papers) and ISI citations (62 996) and constructed the citation distribution for all his papers up to the present date independently of the publication year.

In collecting the statistics for MEDLINE/PubMEd articles we have used the trend for the total number of publications from [41], and the yearly total number of citations statistics from [42].

For Facebook we registered as a developer, and as such we could download all relevant information for the posts of selected users. We have selected 16 popular Facebook pages with users of different background: New York Times, CNN, BBC news, NASA, National Geography, Cristiano Ronaldo, Burning Man, Sugar Factory, USA gov., European Council, IMDB, European Parliament, Democratic Party of USA, Republican Party of USA. From their metadata we have extracted the number of "shares" for all posts independently of their publication date. We have also combined all "share" numbers for the posts of all 16 users and considered as the combined FB database.

Probability distribution functions were constructed using a logarithmic binning method, considering bins of sizes $2^n$. In order not to overload Figure 1 we have plotted the results only for some selected datasets (see the legend). The other collected data, follows the same general trend. All the rescaled data can be nicely fitted with a TPo distribution with $g \approx 1.4$.

**Author Contributions**
Z.N. conceived the study, interpreted the data and wrote the base version of the paper, L.V. collected and elaborated the data and T.B proposed the specific master-equation approach in this paper. All authors have contributed to the final version of the manuscript.

**Acknowledgments**
Work supported by a STAR-UBB fellowship.

**Competing Interests**
The authors declare no competing interests.

**Appendix 1**

We consider a classical master equation approach for the growth phenomenon. This approach is the simplest possible mean-field like description where the properties of different elements (posts, publications) are coarse-grained and only the stochastic character of the process is kept. In this framework, the stochastic growth process is quantified by a mean growth rate $\mu_n$ describing the transition rate from state with $n$ quanta (citations, shares, likes…) to a state with $n+1$ quanta. Since there is no reverse process inside the chain, just a continuous growth a detailed balance condition cannot be fulfilled. We illustrate this process in the left panel of Figure 3, where $N_n(t)$ denotes the number of elements having $n$ quanta at time moment $t$. A master equation for this process writes as:

$$\frac{dN_n(t)}{dt} = \mu_{n-1} N_{n-1}(t) - \mu_n N_n(t) \qquad (1.1)$$

Parallel with this continuous growth there is however a continuous dilution in the system, since the number of elements are continuously increasing in time. This means that

$$N(t) = \sum_n N_n(t) \qquad (1.2)$$

is increasing in time. Considering now the probability $P_n(t)$ that an element has $n$ quanta at time moment $t$

$$P_n(t) = \frac{N_n(t)}{N} \quad , \qquad (1.3)$$

we rewrite the master equation using instead of $N_n(t)$ the $P_n(t)$ distribution:

$$\frac{d}{dt}(NP_n) = N\frac{dP_n}{dt} + P_n \frac{dN_n}{dt} = \mu_{n-1} NP_{n-1} - \mu_n NP_n \qquad (1.4)$$

The number of elements in the systems considered in this work are exponentially increasing. It is well-known that the number of scientific publications is exponentially increasing [34,35]. From a recent statement form Mark Zuckenberg [36] we also learn that the total sharing activity on Facebook is also exponentially growing. This increase is partly due to the fact that the number of Facebook users are growing in time and the average time spent by a user online is also growing. Assuming thus an exponential growth in $N(t)$ with a rate $\gamma$ that is characteristic for the considered ensemble (scientific papers, Facebook posts, etc…):

$$N(t) = N(0)e^{\gamma t} \rightarrow \frac{dN(t)}{dt} = \gamma N(t) \qquad (1.5)$$

from equation (1.4) we arrive to a master equation in $P_n(t)$:

$$\frac{dP_n}{dt} = \mu_{n-1} P_{n-1} - (\mu_n + \gamma) P_n \qquad (1.6)$$

The flow diagram for this process is illustrated in the right panel of Figure 3. The corresponding equation for the $n=0$ term can be obtained from the normalization condition $\sum_n P_n(t) = 1$:

$$\frac{dP_0}{dt} = \gamma - (\mu_0 + \gamma) P_0 \qquad (1.7)$$

We can consider now the continuous limit of equation (1.6) (see for example [4]), where the discrete states $n$ are replaced by continuous $x$ states:

$$\frac{\partial P(x,t)}{\partial t} = -\frac{\partial}{\partial x}\left(\mu(x) P(x,t)\right) - \gamma P(x,t) + \gamma \delta(x) \qquad (1.8)$$

This equation describes a flow with a general velocity field $\mu(x)$, a loss rate $\gamma$ and a feeding at $x=0$. (We denoted by $\delta(x)$ the Dirac functional)

The $P_s(x)$ stationary probability density can be derived from the condition: $\frac{\partial P_s(x,t)}{\partial t} = 0$, and according to (1.8) it satisfies

$$\frac{d}{dx}\left(\mu(x)P_s(x)\right) = -\gamma P_s(x). \qquad (1.9)$$

The solution of this equation writes as

$$P_s(x) = \frac{K}{\mu(x)} e^{-\gamma \int \frac{1}{\mu(x)} dx} \qquad (1.10)$$

In order to write up the solution one has to specify a kernel for the $\mu(x)$ growth rate. From several social-economic phenomena we learn that the growth is usually governed by a preferential selection, in the simplest case by a linear preferential growth rate (the well-known "rich gets richer" phenomenon or the Matthew effect [37,38]), according to which:

$$\mu(x) = \sigma \cdot (x+b) \qquad (1.11)$$

where the $\sigma$ and $b$ values are characteristic to the considered group (scientist, Facebook users).

Accepting this kernel, equation (10) leads us to the Tsallis-Pareto distribution [20,21]:

$$P_s(x) = \frac{\gamma}{b\sigma}\left(1 + \frac{x}{b}\right)^{-1-\gamma/\sigma} \qquad (1.12)$$

Denoting $g = \gamma/\sigma$ and using $b = \langle x \rangle (g-1)$, where $\langle x \rangle$ is the first moment of the distribution, we get:

$$P_s(x) = \frac{g}{(g-1)\langle x \rangle}\left(1 + \frac{x}{(g-1)\langle x \rangle}\right)^{-1-g} \qquad (1.13)$$

This is the scaling Tsallis-Pareto distribution, which for $g = 1.4$ offers a good fit for all the collapsed data on Figure 1. The prediction of our simple model is in agreement with the more technical approach considered in [31].

From this simple mean-field type model we learn that the popularity measures both for scientific publications and Facebook are the results of an exponential growth and a preferential retransmission of the received information. The collapse for the Facebook popularity measures and scientific citations indicate that for their coarse-grained dynamics the ratio $g = \gamma/\sigma$ should be similar. Seemingly this ratio is also independent on the precise manner in how we construct the ensembles (institutes, journals, individuals, etc…). This is an exciting finding which inspires further studies.

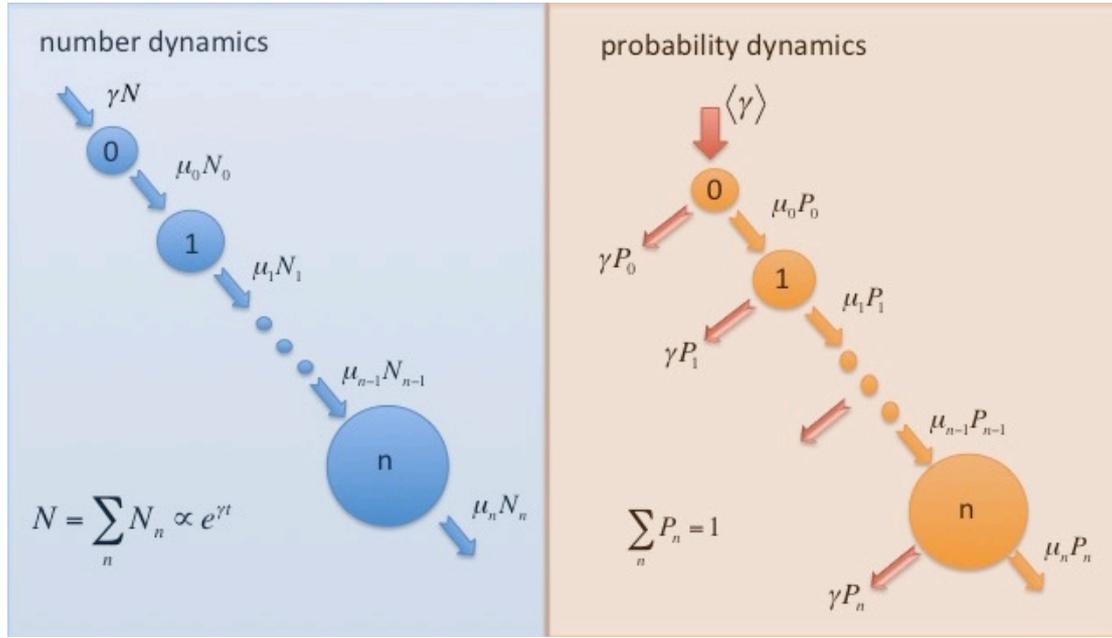

**Figure 3**. Schematic representation of the coarse-grained random growth model considered in the model. The panel on the left side indicates the growth process in the number of elements with $n$ quanta: $N_n$. Due to the fact that the total number of elements is exponentially increasing, the probability $P_n$ that an element will have $n$ quanta, experiences the dynamics sketched on the right panel of the figure.

**Appendix 2. Trend for the average number of citations per paper**

The total number of citations at time $t$ can be written as: $C(t) = \sum_n n N_n(t)$ According to our hypothesis (see Appendix 1) the increase in the total number of citations in unit time is given as: $\dfrac{dC(t)}{dt} = \sum_n \dfrac{dN_n^+}{dt} = \sum_n \sigma(n+b)N_n = \sigma C(t) + \sigma b N(t)$ Combining this with the exponential growth of $N(t)$: $\dfrac{dN(t)}{dt} = \gamma N(t)$ leads to a simple differential equation for the $m(t) = C(t)/N(t)$ average number of citations per work: $\dfrac{dm(t)}{dt} = (\sigma - \gamma)m(t) + \sigma b$ . The solution is an exponential relaxation:

$$m(t) = \dfrac{K}{(\gamma - \sigma)} e^{-(\gamma - \sigma)t} + \dfrac{b}{g-1}, \quad (2.1)$$

where K is an integration constant. Due to the fact that $g = \gamma/\sigma \approx 1.4$ we get $\gamma > \sigma$ and therefore $c(t)$ has an exponentially relaxing trend. From the Supplementary material we learn that $b/(g-1) = \langle m \rangle_{ech}$, which is the equilibrium value for the average citation per paper in the considered ensemble.

We can now determine the time-evolution of the total citations number per year. Let us assume now that we measure the time in years, and introduce the yearly published article number $n(t) = dN(t)/dt = n_0 \exp[-\gamma(t-t_0)]$ , and the new citations that appear in one year: $c(t) = dC(t)/dt$. If we assume that at time $t_0$ we have $c(t_0) = c_0$ and $n(t_0) = n_0$ we get that

$$K = \left[\frac{c_0}{n_0}(g-1) - b\right]\frac{\gamma}{g}, \quad (2.2)$$

and

$$c(t) = c_0 e^{\sigma(t-t_0)} + \frac{bn_0}{(g-1)}\left(e^{\gamma(t-t_0)} - e^{\sigma(t-t_0)}\right). \quad (2.3)$$